% -------------------------------------------------------------------------
%    nig   Feb 1998; 20 May 1998 ;  18 Dec 98  With graphics
% -------------------------------------------------------------------------
\input psfig.sty
%-------------------------------------

\def\ptitle{Constructive Inversion of Energy Trajectories in Quantum Mechanics}
\nopagenumbers
\magnification=\magstep1
\hsize 6.0 true in  \hoffset 0.25 true in         % 6 in width with 1.25 in margins default = (6.5, 0)
\emergencystretch=0.6 in                            % TEXBook p 107 : allows h-space 
\vfuzz 0.4 in                                                % page-length flexibility
\hfuzz  0.4 in                                               % line-length flexibility
\mathsurround=2pt                                     % Default is 2pt
\topskip=24pt                                             % Default is 10pt
\lineskip=3pt                                              % default is 1pt
% --------------------------------------------------------------------
%  generic unix fonts (lower case names)
% --------------------------------------------------------------------
\font\tr=cmr10                          % Our default
\font\bf=cmbx10                         % Redefinition
                         % Redefinition
\font\it=cmti10                         % Redefinition
\font\trbig=cmbx10 scaled 1500          % Main Title
                          % Theorems                       
\font\tiny=cmr8                         % Running title
% --------------------------------------------------------------------

                                    % New line
\def\title#1{\bigskip\noindent\bf #1 ~ \tr\smallskip} % Headings
                     % Math Sets eg R -> |R
\def\mb#1{\hbox{\bf#1}}                          % bold in math mode
                         % small bold in math mode
\def\ng{>\kern -9pt|\kern 9pt}                   % not greater than
                  % bra ket:  math mode (to replace angle)
                                       %   ket  >
\def\hi#1#2{$#1$\kern -2pt-#2}                % hyphen  eg      \hi{N}{body} = N-body
\def\hy#1#2{#1-\kern -2pt$#2$}               % hyphen eg       hy{large}{N} = large-N

\def\half{{1 \over 2}}

\output={\shipout\vbox{\makeheadline
                                      \ifnum\the\pageno>1 {\hrule}  \fi 
                                      {\pagebody}   
                                      \makefootline}
                   \advancepageno}

\headline{\noindent {\ifnum\the\pageno>1 
                                   {\tiny \ptitle} \fi}\hfil {\tiny page~\the\pageno}}
\footline{}

  %  QED
 % SQUARE 

% -------------------------------------------------------------------------------- ref.tex
\newcount\zz  \zz=0  % switch for printing references
\newcount\q   %  reference number
\newcount\qq    \qq=0  % starting reference number-1   (usually zero)

\def\pref #1#2#3#4#5{\frenchspacing \global \advance \q by 1     % paper reference
    \edef#1{\the\q}
       {\ifnum \zz=1 { %
         \item{[\the\q]} 
         {#2} {\bf #3},{ #4.}{~#5}\medskip} \fi}}

\def\bref #1#2#3#4#5{\frenchspacing \global \advance \q by 1     % book reference
    \edef#1{\the\q}
    {\ifnum \zz=1 { %
       \item{[\the\q]} 
       {#2}, {\it #3} {(#4).}{~#5}\medskip} \fi}}

\def\gref #1#2{\frenchspacing \global \advance \q by 1             % general reference
    \edef#1{\the\q}
    {\ifnum \zz=1 { %
       \item{[\the\q]} 
       {#2}\medskip} \fi}}

 \def\sref #1{~[#1]}

\def\references#1{\zz=#1
   \parskip=2pt plus 1pt   % default is 0pt plus 1pt       
   {\ifnum \zz=1 {\noindent \bf References \medskip} \fi} \q=\qq

  \bref{\chad}{K. Chadan and P. C. Sabatier}{Inverse Problems in Quantum    Scattering Theory}{Springer, New York, 1989}{The `inverse problem in the coupling constant' is discussed on p406}
\pref{\halla}{R. L. Hall, Phys. Rev A}{51}{1787 (1995)}{}          % WKB
\pref{\hallb}{R. L. Hall, J. Phys. A:Math. Gen}{25}{4459 (1992)}{} % refining theorem
\pref{\hallc}{R. L. Hall, Phys. Rev. A}{50}{2876 (1995)}{}         % flat bottoms
\pref{\halld}{R. L. Hall, J. Phys. A:Math. Gen}{28}{1771 (1995)}{} % geom spect inversion
\bref{\raab}{O. L. De Lange and R. E. Raab}{Operator Methods in Quantum Mechanics}{Oxford University Press, Oxford, 1991}{Perturbed harmonic oscillators with identical spectra to that generated by $f(x) = x^{2}$ are given on p71}
\pref{\mass}{H. S. W. Massey and C. B. O. Mohr, Proc. Roy. Soc.}{148}{206 (1934)}{}
\bref{\flug}{S. Fl\"ugge}{Practical Quantum Mechanics}{Springer, New York, 1974}{The exponential potential is discussed on p196 and the sech-squared potential on p94}
\pref{\raba}{P. Gross, H. Singh, and H. Rabitz, Phys. Rev. A}{47}{4593 (1993)}{}
\pref{\rabb}{Z. Lu and H. Rabitz, Phys. Rev. A}{52}{1961 (1995)}{}

 }

 \references{0}    % Initialization of reference numbers
% -------------------------------------------------------------------------------- ref.tex 

% --------------------------------------------------------------------------------- 
%   Title page
% ---------------------------------------------------------------------------------
\vglue 0.1true in
\parskip=15pt plus 1pt   % default is 0pt plus 1pt : 15pt is only for the title page
\baselineskip = 16 true pt
\vskip 1.0true in
\centerline{\trbig Constructive Inversion of Energy Trajectories}
\vskip 0.1true in
\centerline{\trbig in Quantum Mechanics}
\vskip 0.5true in
\tr                                                                  % use our default font
\centerline{Richard L. Hall}\medskip
\centerline{Department of Mathematics and Statistics,}
\centerline{Concordia University,}
\centerline{1455 de Maisonneuve Boulevard West,}
\centerline{Montr\'eal, Qu\'ebec,}
\centerline{Canada H3G 1M8.}
\bigskip\bigskip\bigskip
% \item{Tel}~~(514) 848-3221 or 848-3222
%\item{Fax}~~(514) 848-4511
%\item{E-Mail}~~rhall@cicma.concordia.ca

\vglue 0.1true in

\centerline{\bf Abstract}

We suppose that the ground-state eigenvalue $E = F(v)$ of the Schr\"odinger Hamiltonian $H = -\Delta + vf(x)$ in one dimension is known for all values of the coupling $v > 0.$   The potential shape $f(x)$ is assumed to be symmetric, bounded below, and monotone increasing for $x > 0.$ A fast algorithm is devised which allows the potential shape $f(x)$ to be  reconstructed from the energy trajectory $F(v).$  Three examples are discussed in detail: a shifted power-potential, the exponential potential, and the sech-squared potential are each reconstructed from their known exact energy trajectories. 
\item{PACS}~~03 65 Ge 

\vfil\break
\parskip=5pt plus 1pt                                                  % MAIN PARSKIP
\baselineskip = 16true pt                                             % baselineskip
% ---------------------------------------------------------------------------------------
    \title{1.~~Introduction}
% ---------------------------------------------------------------------------------------

This paper is concerned with what may be called `geometric spectral inversion'.  We suppose that a discrete eigenvalue $E = F(v)$ of the Schr\"odinger Hamiltonian
$$H = -\Delta + vf(x)\eqno{(1.1)}$$
is known for all sufficiently large values of the coupling parameter $v > 0$ and we try to use this data to reconstruct the potential shape $f.$  The usual `forward' problem would be: given the potential (shape) $f(x),$ find the energy trajectory $F(v);$ the problem we now consider is the inverse of this $F \rightarrow f.$

This problem must at once be distinguished from the `inverse problem in the coupling constant' discussed, for example, by Chadan and Sabatier\sref{\chad}. In this latter problem, the discrete part of the `input data' is a set $\{v_{i}\}$ of values of the coupling constant that all yield the identical energy eigenvalue $E.$ The index $i$ might typically represent the number of nodes in the corresponding eigenfunction.  In contrast, for the problem discussed in the present paper, $i$ is kept fixed and the input data is the graph $(F(v),v),$ where the coupling parameter has any value $v > v_c,$ and $v_c$ is the critical value of $v$ for the support of a discrete eigenvalue with $i$ nodes.  We shall mainly discuss the bottom of the spectrum $i = 0$ in this paper. However, on the basis of results we have obtained for the inversion IWKB of the WKB approximation\sref{\halla}, there is good reason to believe that constructive inversion may also be possible starting from any discrete eigenvalue trajectory $F_{i}(v),$ $i > 0.$  In fact, perhaps not surprisingly, IWKB yields better results starting from higher trajectories; moreover, they become asymptotically exact as the eigenvalue index is increased without limit.  

By making suitable assumptions concerning the class of potential shapes, theoretical progress has already been made with this inversion problem\sref{\hallb-\halld}.  The most important assumptions that we retain throughout the present paper are that $f(x)$ is symmetric, monotone increasing for $x >0,$ and bounded below: consequently the minimum value is $f(0).$  We assume that our spectral data, the energy trajectory $F(v),$ derives from a potential shape $f(x)$ with these features.  We have discussed\sref{\hallb} how two potential shapes $f_1$ and $f_2$ can cross over and still preserve spectral ordering $F_1 < F_2.$  It is known\sref{\hallc} that lowest point $f(0)$ of $f$ is given by the limit
$$f(0) = {\lim_{v\rightarrow\infty}} {F(v) \over v}.\eqno{(1.2)}$$
We have proved\sref{\hallc} that a potential shape $f$ has a finite flat portion ($f'(x) = 0$) in its graph starting at $x = 0$ if and only if the mean kinetic energy is bounded.  That is to say, $s = F(v) - vF'(v) < K,$ for some positive number $K.$  More specifically, the size $b$ of this patch can be estimated from $F$ by means of the inequality:
$$s \leq K \quad \Rightarrow\quad f(x) = f(0),\quad |x| \leq b, \quad {\rm and} \quad b = {\pi \over 2} K^{-{1 \over 2}}.\eqno{(1.3)}$$
The monotonicity of the potential, which allows us to prove results like this, also yields the\hfil\break 
\noindent{\bf Concentration Lemma}\sref{\hallc}
$$q(v) = \int_{-a}^{a}\psi^2(x,v)dx > {{f(a) - F'(v)} \over {f(a) - f(0)}} \quad \rightarrow \quad 1, \quad v \rightarrow \infty,\eqno{(1.4)}$$
where $\psi(x,v)$ is the normalized eigenfunction satisfying $H\psi = F(v)\psi.$  More importantly, perhaps, if $F(v)$ derives from a symmetric monotone potential shape $f$ which is bounded below, then $f$ is {\it uniquely} determined\sref{\halld}. The significance of this result can be appreciated more clearly upon consideration of an example. Suppose the bottom of the spectrum of $H$ is given by $F(v) = \surd{v},$ what is $f?$  It is well known, of course, that $f(x) = x^{2} \rightarrow F_{o}(v) = \surd{v};$ but are there any others?  Are scaling arguments reversible?  A possible source of disquiet for anyone who ponders such questions is the uncountable number of (unsymmetric) perturbations\sref{\raab} of the harmonic oscillator all of which have the identical spectrum to that of the unperturbed oscillator $f(x) = x^{2}$.

If, in addition to symmetry and monotonicity, we also assume that a potential shape $f_{1}(x)$ vanishes at infinity and that $f_{1}(x)$ has area,  then a given trajectory function $F_{1}(v)$ corresponding to $f_{1}(x)$ can be `scaled'\sref{\halld} to a standard form in which the new function $F(v) = \alpha F_{1}(\beta v)$ corresponds to a potential shape $f(x)$ with area $-2$ and minimum value $f(0) = -1.$  Thus square-well potentials, which of course are completely determined by depth and area, are immediately invertible; moreover it is known that,  amongst all standard potentials, the square-well it `extremal' for it has the lowest possible energy trajectory.  In Ref.[\halld] an approximate variational inversion method is developed; it is also demonstrated constructively that all separable potentials are invertible.  However, these results and additional constraints are not used in the present paper.  When a potential has area $2A$, we first assumed, during our early attempts at numerical inversion, that it would be very useful to determine $A$ from $F(v)$ and then appropriately constrain the inversion process. However, the area constraint did not turn out to be useful.  Thus the numerical method we have established for constructing $f(x)$ from $F(v)$ does not depend on use of this constraint, and is therefore not limited to the reconstruction of potentials which vanish at infinity and have area.
 
Much of numerical analysis assumes that errors arising from arithmetic computations or from the computation of elementary functions is negligibly small.  The errors usually studied in depth are those that arise from the discrete representation of continuous objects such as functions, or from operations on them, such as derivatives or integrals.  In this paper we shall take this separation of numerical problems to a higher level.  We shall assume that we have a numerical method for solving the eigenvalue problem in the {\it forward} direction $f(x) \rightarrow F(v)$ that is reliable and may be considered for our purposes to be essentially error free.  Our main emphasis will be on the design of an effective algorithm for the inverse problem {\it assuming} that the forward problem is numerically soluble.  The forward problem is essential to our methods because we shall need to know not only the given exact energy trajectory $F(v)$ but also, at each stage of the reconstruction, what eigenvalue a partly reconstructed potential generates. This line of thought immediately indicates that we shall also need a way of temporarily extrapolating a partly reconstructed potential to all $x.$

Our constructive inversion algorithm hinges on the assumed symmetry and monotonicity of $f(x).$ This allows us to start the reconstruction of $f(x)$ at $x = 0,$ and sequentially increase x. In Section (2) it is shown how numerical estimates can be made for the shape of the potential near $x = 0,$ that is for $x < b,$ where $b$ is a parameter of the algorithm. In Section (3) we explore the implications of the potential's monotonicity for the `tail' of the wave function.   In Section (4) we establish a numerical representation for the form of the unknown potential for $x > b$ and construct our inversion algorithm.  In Section (5) the algorithm is applied to three test problems.       

% -----------------------------------------------------------
    \title{2.~~The reconstruction of $f(x)$ near $x = 0.$}
% -----------------------------------------------------------
Since the energy trajectory $F(v)$ which we are given is assumed to arise from a symmetric monotone potential, and since the spectrum generated by the potential is invariant under shifts along the \hi{x}{axis}, we may assume without loss of generality that the minimum value of the potential occurs at $x = 0.$  We now investigate the behaviour of $F(v),$ either analytically or numerically, for large values of $v.$  The purpose is to establish a value for the starting point $x = b > 0$ of our inversion algorithm and the shape of the potential in the interval $x \in [0,b].$  First of all, the minimum value $f(0)$ of the potential is provided by the limit (1.2).  Now, if the mean kinetic energy $s = (\psi, -\Delta \psi) = F(v) - vF'(v)$ is found to be bounded above by a positive number $K,$ then we know\sref{\hallc} that the potential shape $f(x)$ satisfies $f(x) = f(0),\quad x \in [0,b],$ where b is given by (1.3).  In this case we have a value for $b$ and also the shape $f(x)$ inside the interval $[0,b].$

If the mean potential energy $s$ is (or appears numerically to be) unbounded, then we adopt another strategy: we model $f(x)$ as a shifted power potential near $x = 0.$  Since we never know $f(x)$ {\it exactly,} we shall need another symbol for the approximation we are currently using for $f(x).$   We choose this to be $g(x)$ and we suppose that the bottom of the spectrum of $-\Delta + vg(x)$ is given by $G(v).$  The goal is to adjust $g(x)$ until $G(v)$ is close to the given $F(v).$  Thus we write
$$f(x) \approx g(x) = f(0) + Ax^{q},\quad x \in [0,b].\eqno{(2.1)}$$
Therefore we have three positive parameters to determine, $b,\ A,$ and $q.$  We first suppose  that $g(x)$ has the form (2.1) for {\it all} $x \geq 0.$  We now choose a `large' value $v_{1}$ of $v.$  This is related to the later choice of $b$ by a bootstrap argument: the idea is that we choose $v_{1}$ so large that the turning point determined by 
$$\psi_{xx}(x,v_1)/\psi(x,v_1) = v_{1}f(x) - F(v) = 0\eqno{(2.2)}$$
is equal to $b.$  The concentration lemma guarantees that this is possible.  By scaling arguments we have 
$$G(v) = f(0)v + E(q)(vA)^{2 \over {2+q}},\eqno{(2.3)}$$
where $E(q)$ is the bottom of the spectrum of the pure-power Hamiltonian $-\Delta + |x|^{q}.$  We now `fit' $G(v)$ to $F(v)$ by the equations $G(v_{1}) = F(v_{1})$ and $G(2v_{1}) = F(2v_{1})$  which yield the estimate for $q$ given by
$$\eta = {2 \over {2 + q}} = {{\log(F(2v_{1})-2v_{1}f(0)) - \log(F(v_{1})-v_{1}f(0))} \over {\log(2)}}.\eqno{(2.4)}$$
Thus $A$ is given by
$$A = \left((F(v_{1}) - v_{1}f(0))/E(q)\right)^{1 \over \eta}/v_{1}.\eqno{(2.5)}$$
We choose $b$ to be equal to the turning point corresponding to the {\it model} potential $g(x)$ with the smaller value of $v,$ that is to say so that $f(0) + Ab^{q} = F(v_{1})/v_{1},$ or
$$b = \left({{F(v_{1})-v_{1}f(0)} \over {Av_{1}}}\right)^{1 \over q}.\eqno{(2.5)}$$
Thus we have determined the three parameters which define the potential model $g(x)$ for $x \in [-b,b].$  
% --------------------------------------------------------
    \title{3.~~The tail of the wavefunction}
% --------------------------------------------------------
Let us suppose that the ground-state wave function is $\psi(x,v).$ Thus the turning point $\psi_{xx}(x,v) = 0$ occurs for a given $v$ when 
$$x = x_{t}(v) = f^{-1}(R(v)),\quad R(v) = \left({{F(v)} \over v}\right).\eqno{(3.1)}$$
The concentration lemma (1.4) quantifies the tendency of the wave function to become, as the coupling $v$ is increased, progressively more concentrated on the patch $[-c,c],$ where $x = c$ is the point (perhaps zero) where $f(x)$ first starts to increase.  This allows us to think in terms of the wave function having a `tail'.  We think of a symmetric potential as having been determined from $x = 0$ up to the current point $x.$  The question we now ask is: what value of $v$ should we use to determine how $f(x),$ or, more particularly, our {\it approximation} $g(x)$ for $f(x),$ continues beyond the current point.  We have found that a good choice is to choose $v$ so that the turning point $x_{t}(v) = x/2,$ or some other similar fixed fraction $\sigma < 1$ of the current $x$ value. The algorithm seems to be insensitive to this choice.  Since $g(x)$ has been constructed up to the current point, and $F(v)$ is known, the value of v required follows by inverting (3.1).  It has been proved\sref{\hallc} that R(v) is monotone and therefore invertible.  Hence we have the following general recipe for $v:$ 
$$v = R^{-1}(g(\sigma x)),\quad \sigma = {1 \over 2}.\eqno{(3.2)}$$
Since we can only determine Schr\"odinger eigenvalues of $H = -\Delta + vg(x)$ if the potential is defined for all $x,$ we must have a policy about temporarily extending $g(x).$   We have tried many possibilities and found the simplest and most effective method is to extend $g(x)$ in a straight line, with slope to be determined.

In Figure~(1) we illustrate the ideas just discussed for the case of the sech-squared potential.  The inset graph shows the sech-squared potential perturbed from $x = x_{a}$ by five straight line extensions; meanwhile the main graph shows the corresponding set of five wave functions which agree for $0 \leq x \leq x_{a}$ and then continue with different `tails' dictated by the corresponding potential extensions.  The value of the coupling $v$ is the value that makes the turning point of the wave function occur at $x = x_{a}/2.$  This figure illustrates the sort of graphical study that has lead to the algorithm described in this paper. 

% -------------------------------------------------------
      \title{4.~~The inversion algorithm}
% -------------------------------------------------------

We must first define the `current' approximation $g(x)$ for the potential $f(x)$ sought. For values of $x$ less than $b,$ $g(x)$ is defined either as the horizontal line $f(x) = f(0)$ or as the shifted power potential (2.1).  For values of $x$ greater than $b,$ the \hi{x}{axis} is divided into steps of length $h.$ Thus the `current' value of $x$ would be of the form $x = x_{k} = b + kh,$ where $k$ is a positive integer.  The idea is that $g(x_{k})$ is determined sequentially and $g(x)$ is interpolated linearly between the $x_{k}$ points.  We suppose that $\lbrace g(x_{k})\rbrace$ have already been determined up to $k$ and we need to find $y = g(x_{k+1}).$  For $x \geq x_{k}$ we let
$$g(x) = g(x_{k}) + (y - g(x_{k})) {{x - x_{k}} \over h}.\eqno{(4.1)}$$
If, from a study of $F(v),$ the underlying potential $f(x)$ has been shown\sref{\halld} to be bounded above, it is convenient to rescale $F(v)$ so that it corresponds to a potential shape $f(x)$ which vanishes at infinity.  In this case  it is slightly more efficient to modify (4.1) so that for large $x$ the straight-line extrapolation of $g(x)$ is `cut' to zero instead of becoming positive.  In either case we now have for the current point $x_{k}$ an approximate potential $g(x)$ parameterized by the `next' value $y = g(x_{k+1}).$  The task of the inversion algorithm is simply to choose this value of $y.$

Let us suppose that, for given values of $k$ and $y,$ the bottom of the spectrum of $H = -\Delta + vg(x)$ is given by $G(v,k,y),$ then the inversion algorithm may be stated in the following succinct form in which $\sigma < 1$ is a fixed parameter.  Find $y$ such that 
$$vg(\sigma x_{k}) = F(v) = G(v,k,y);\quad {\rm then}\quad g(x_{k+1}) = y.\eqno{(4.2)}$$
The value of $v$ is first chosen so that the turning point of the wave function generated by $g$ occurs at $\sigma x_{k};$ after this, the value of $y$ is chosen so that $G$ `fits' $F$ for this value of $v.$  The value of the parameter $\sigma$ chosen for the examples discussed in section~(5) below is $\sigma = {1 \over 2}.$   The idea behind this choice can best be understood from a study of Figure~(1): the value of the coupling $v$ must be such that the current value of $x$ for which $y$ is sought is in the `tail' of the corresponding wave function; that is to say, the turning point $\sigma x$ should be before $x,$ but not too far away.  Fortunately the inversion algorithm seems to be insensitive to the choice of $\sigma.$
\hfil\vfil\break %%%%%%%%%%%%%%%%%%%%%%%%%%%%%%%%%%%%%%%%%%%%%%%%%%%%KLUDGE BREAK %%%%%%%%%%  
% -------------------------------------------------------
      \title{5.~~Three examples}
% -------------------------------------------------------
The first example we consider is the unbounded potential whose shape $f(x)$ and corresponding exact energy trajectory $F(v)$ are given by the $\{f,F\}$ pair
$$f(x) = -1 + |x|^{3 \over 2}\quad \longleftrightarrow \quad F(v) = -v + E(3/2)v^{4 \over 7},\eqno{(5.1)}$$
where $E(3/2)$ is the bottom of the spectrum of $H = -\Delta + |x|^{3 \over 2}$ and has the approximate value $E(3/2) \approx 1.001184.$  Applying the inversion algorithm to $F(v)$ we obtain the reconstructed potential shown in Figure~(2).  We first set $v_1 = 10^4$ and find that the initial shape is determined (as described in Section~(2)) to be $-1 + x^{1.5}$ for $x < b = 0.072.$ For larger values of $x$ the step size is chosen to be $h = 0.05$ and $40$ iterations are performed by the inversion algorithm.  The results are plotted as hexagons on top of the exact potential shape shown as a smooth curve.  This entire computation takes less than $20$ seconds with a program written in C++ running on a $200$MHz Pentium Pro.

The following two examples are bounded potentials both having \hy{large}{x} limit zero, lowest point $f(0) = -1,$ and area $-2.$ The exponential potential\sref{\mass,\flug} has the $\{f,F\}$ pair
$$f(x) = -e^{-|x|}\quad \longleftrightarrow \quad J'_{2|E|^{1 \over 2}}(2v^{1 \over 2}) = 0 \quad \equiv\quad E = F(v),\eqno{(5.2)}$$
where $J'_{\nu}(x)$ is the derivative of the Bessel function of the first kind of order $\nu.$  For the sech-squared potential\sref{\flug} we have 
$$f(x) = -{\rm sech}^{2}(x)\quad \longleftrightarrow \quad F(v) = -\left[\left(v+{1 \over 4}\right)^{\half} - \half\right]^{2}.\eqno{(5.3)}$$ 
In Figure~(3) the two energy trajectories are plotted.  Since the two {\it potentials} have lowest value $-1$ and area $-2$ it follows\sref{\halld} that the corresponding trajectories both have the form $F(v) \approx -v^{2}$ for small $v$ and they both satisfy the \hy{large}{v} limit $\lim_{v \rightarrow \infty}\left(F(v)/v\right) = -1.$  Thus the differences between the potential shapes is somehow encoded in the fine differences between these two similar energy curves for intermediate values of $v:$ it is the task of our inversion theory to decode this information and reveal the underlying potential shape.  If we apply the inversion algorithm to these two problems we obtain the results shown in Figures~(4) and (5).  The parameters used are exactly the same as for the first problem described above.  The time taken to perform the inversions is again less than $20$ seconds if we discount, in the case of the exponential potential, the extra time taken to compute $F(v)$ itself. 
   
% -------------------------------------------------------
      \title{6.~~Conclusion}
% -------------------------------------------------------
Once we suspect (or know) that an energy trajectory $F(v)$ derives from a potential shape $f(x),$ it is certainly possible in principle to model the potential discretely as $g(x)$ and then find $g$ approximately by a least-squares fit of $G(v)$ to $F(v).$  Such a `brute force' method would not be easy or fast, even for problems in one dimension.  In terms of the reconstructions presented in this paper, one would have to consider minimizing a function of the form $\sum_{i = 1}^{40}|G(v_{i};\mb{Y}) - F(v_{i})|^{2},$ where the vector $\mb{Y}$ represents the $40$ values of $g(x_{k})$ to be determined.  We have found that such a function of $\mb{Y}$ has very erratic behaviour unless the starting point can be chosen quite close to the critical point.     

The purpose of the approach discussed in this paper is however not so much to do with efficiency as with understanding.  The method we have found is intimately linked to the basic properties of the problem: the implications of monotonicity, the relation between the position of the turning point of the wave function and the value of $v,$ and the tail behaviour. The effectiveness of the resulting algorithm stems from its systematic use of all this information.  
If a potential shape $f(x)$ is symmetric but {\it not} monotonic (on the half axis), then for large values of the coupling $v$ the problem will necessarily split into regimes that become more and more isolated as $v$ increases.  The situation could become arbitrarily complicated, perhaps involving resonances, and we have no idea at present whether reconstruction $F\rightarrow f$ would in principle be possible in the general case.

If the potential were unimodal and monotonic away from the minimum point, we do not at present know what might be the spectral inheritance of the additional property of the symmetry of $f(x).$  Is there non-uniqueness in this case?  Could a symmetric potential be constructed that would have the same energy trajectory $F(v)$ as that of a given non-symmetrical unimodal potential shape $f(x)?$  Many interesting questions such as this which are simple to pose nevertheless appear at present to be very difficult to answer.

In our earlier papers on this topic we discussed some suggestions for applications of this form of spectral inversion.  The situations that are most strongly suggestive are those such as the screened-Coulomb potentials used in atomic physics where the coupling varies with the atomic number.  In such a case $F_{n}(v)$ or, more accurately, pair {\it differences} between such functions, would only be known at certain isolated points.  Now that an effective form of constructive inversion is available, it will be possible to consider this more physically important type of application.  Another approach which has not yet been applied to geometric spectral inversion is via control theory.  Rabitz {\it et al}\sref{\raba,\rabb} have successfully used ideas from control theory to reconstruct molecular potentials from sets of data that are directly measurable.  This is the ultimate goal of the present work on geometric spectral inversion.

\title{Acknowledgment}
Partial financial support of this work under Grant No. GP3438 from the Natural Sciences and Engineering Research Council of Canada is gratefully acknowledged.
\hfil\vfil\break
 
\references{1}

\hfil\vfil\break
\hbox{\vbox{\psfig{figure=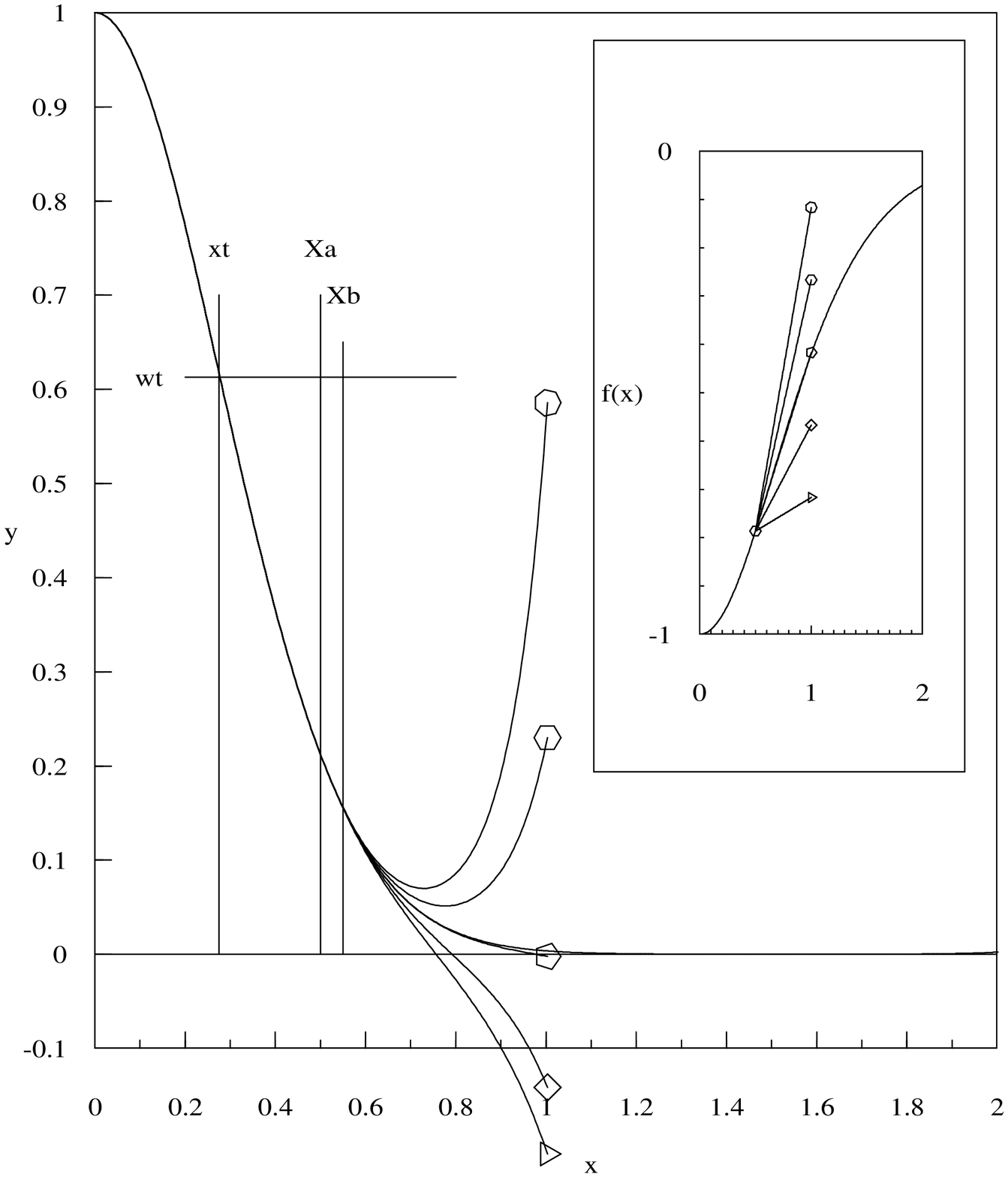,height=6in,width=5in,silent=}}}
\noindent {\bf Figure~(1)}~~The potential $f(x) = -{\rm sech}^{2}(x)$ is perturbed from $x = x_{a}$
 by straight-line segments.  Each segment leads to a perturbation in the tail of the corresponding wave function.
 The coupling $v$ is chosen so that $x_{a} = x_{t}/2,$ where $x_{t}$ is the turning point of the wave function.

\hfil\vfil\break
\hbox{\vbox{\psfig{figure=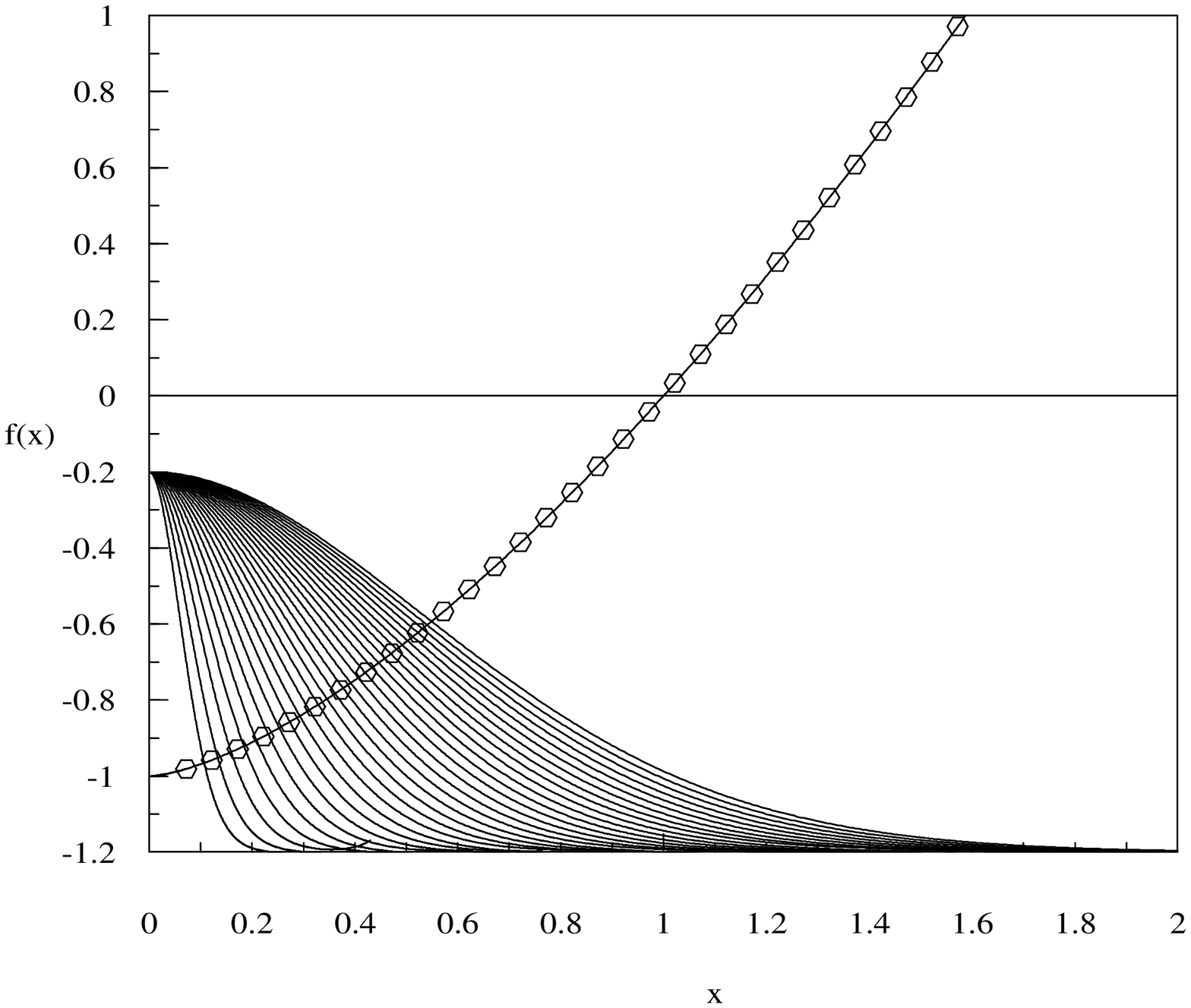,height=6in,width=5in,silent=}}}
\noindent{\bf Figure~(2)}~~Constructive inversion of the energy trajectory $F(v)$ for the shifted
 power potential $f(x) = -1 + |x|^{3 \over 2}.$  For $x \leq b = 0.072,$ the algorithm correctly
 generates the model $f(x);$ for larger values of $x,$ in steps of size $h = 0.05,$ the hexagons
 indicate the reconstructed values for the potential $f(x),$ shown exactly as a smooth curve. 
The unnormalized wave functions are also shown.\medskip

\hfil\vfil\break
\hbox{\vbox{\psfig{figure=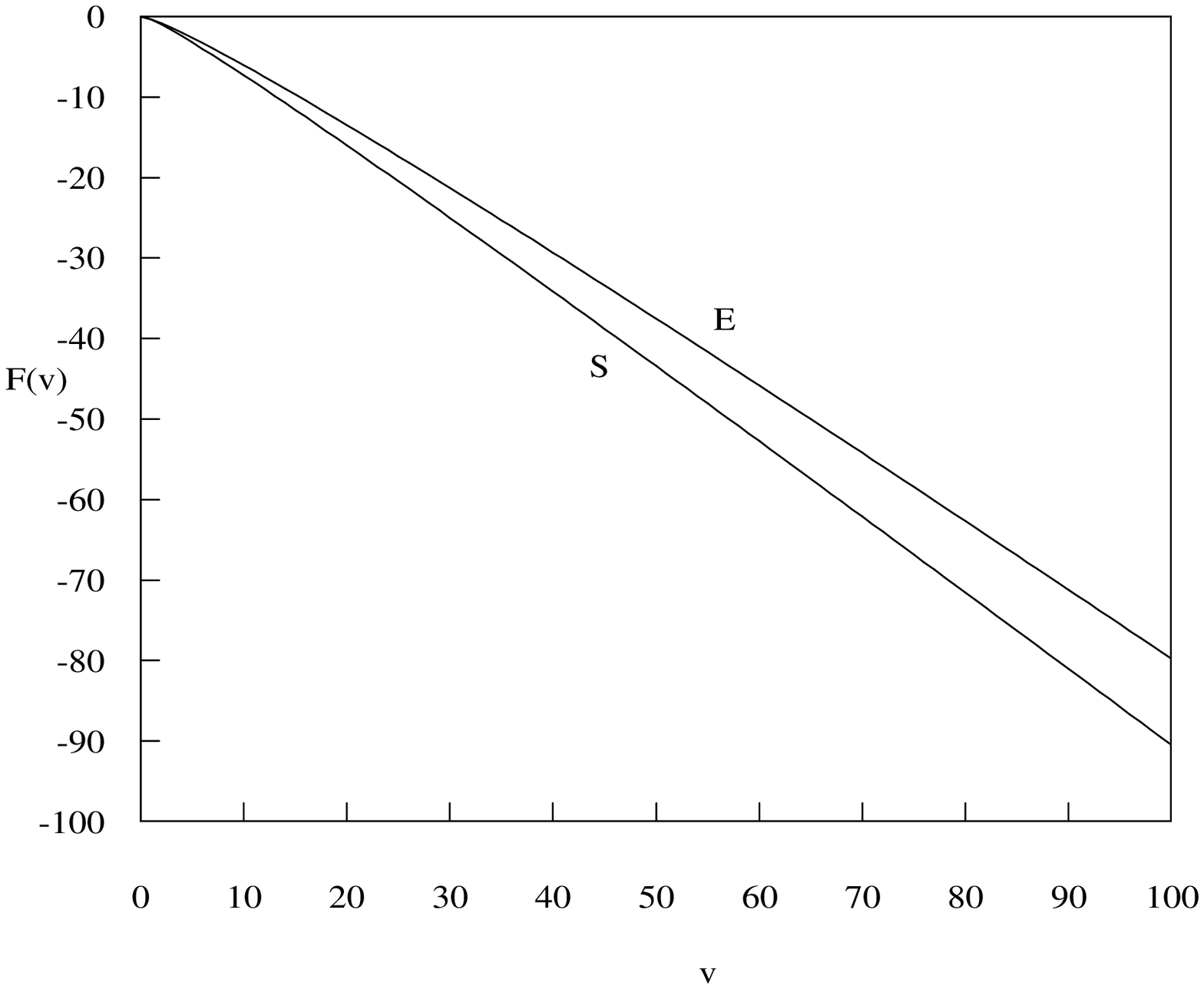,height=6in,width=5in,silent=}}}
\noindent{\bf Figure~(3)}~~The ground-state energy trajectories $F(v)$ for the exponential potential
 (E) and the sech-squared potential (S). For small $v,$ $F(v) \approx -v^{2};
$ for large $v,$ $\lim_{v \rightarrow \infty}\left(F(v)/v\right) = -1.$  The shapes of the underlying
 potentials are buried in the details of $F(v)$ for intermediate values of $v.$\medskip

\hfil\vfil\break
\hbox{\vbox{\psfig{figure=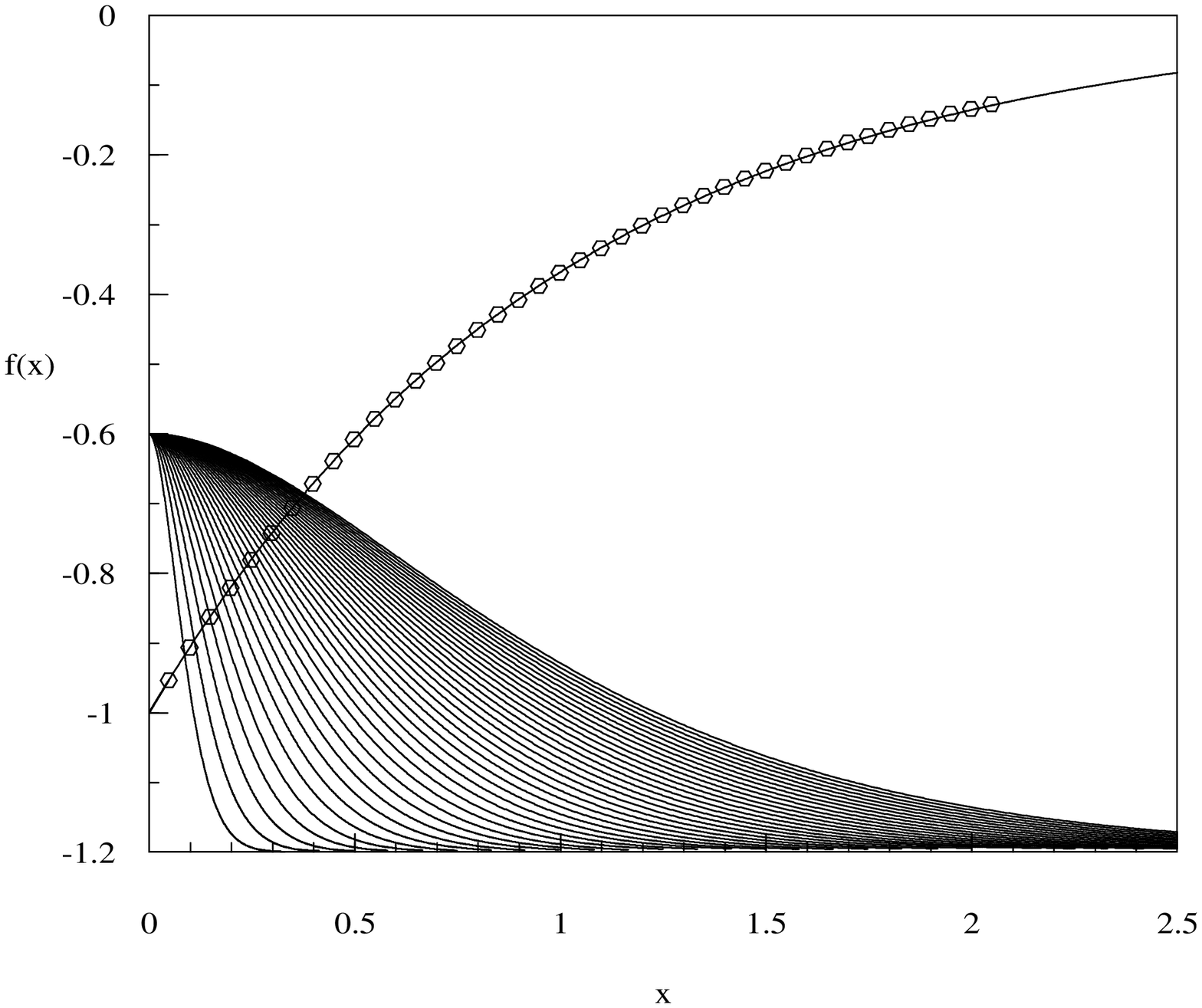,height=6in,width=5in,silent=}}}
\noindent{\bf Figure~(4)}~~Constructive inversion of the energy trajectory $F(v)$ for the
 exponential potential $f(x) = -\exp(x).$  For $x \leq b = 0.048,$ the algorithm correctly
 generates the model $f(x) = -1 + |x|;$ for larger values of $x,$ in steps of size $h = 0.05,$
 the hexagons indicate the reconstructed values for the potential $f(x),$ shown exactly as a
 smooth curve. The unnormalized wave functions are also shown.\medskip

\hfil\vfil\break
\hbox{\vbox{\psfig{figure=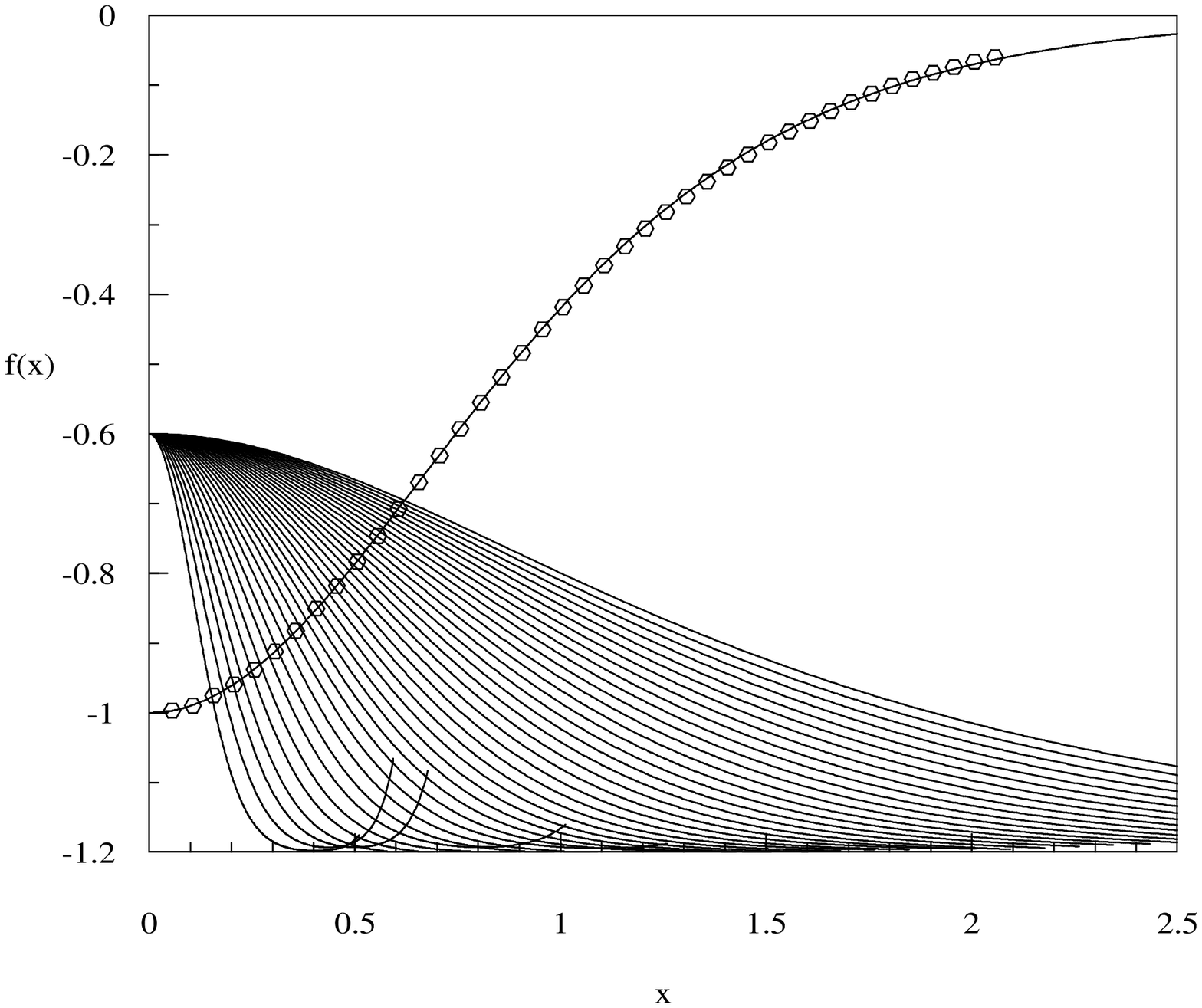,height=6in,width=5in,silent=}}}
\noindent{\bf Figure~(5)}~~Constructive inversion of the energy trajectory $F(v)$ for
 the sech-squared potential $f(x) = -{\rm sech}^{2}(x).$  For $x \leq b = 0.1,$ the
 algorithm correctly generates the model $f(x) = -1 + x^2;$ for larger values of
 $x,$ in steps of size $h = 0.05,$ the hexagons indicate the reconstructed values
 for the potential $f(x),$ shown exactly as a smooth curve. The unnormalized wave functions are also shown.
      
\hfil\vfil
\end